# Proposal of a New Block Cipher reasonably Non-Vulnerable against Cryptanalytic Attacks

**Abhijit Chowdhury[1], Angshu Kumar Sinha[1], Saurabh Dutta[2]**

[1] **Department of Computer Applications, NSHM College of Management and Technology, Durgapur, West Bengal, INDIA**

[2]**Dr. B. C. Roy Engineering College, Durgapur-713206, West Bengal, INDIA**

## Abstract

This paper proposes a new block cipher termed as "Modular Arithmetic based Block Cipher with Varying Key-Spaces (MABCVK)" that uses private key-spaces of varying lengths to encrypt data files. There is a simple but intelligent use of theory of modular arithmetic in the scheme of the cipher. Based on observed implementation of the proposed cipher on a set of real data files of several types, all results are tabulated and analyzed. The schematic strength of the cipher and the freedom of using a long key-space expectedly can make it reasonably non-vulnerable against possible cryptanalytic attacks. As a part of the future scope of the work, it is also intended to formulate and implement an enhanced scheme that will use a carrier image to have a secure transmission of the private key

## 1. Introduction

Module arithmetic [1], a strong backbone in numerous existing ciphering protocols [2], has been applied in an intelligent manner to develop "Modular Arithmetic based Block Cipher with Varying Key-Spaces" with the acronym as MABCVK. Although the encryption of data using MABCVK results in encrypted data of extended dimension, its schematic strength and the liberty of using a long key-space while encrypting using it expectedly can make MABCVK reasonably non-vulnerable against possible cryptanalytic attacks.

Section 2 presents the scheme followed in MABCVK. A report of implementation of MABCVK on a sample plaintext is demonstrated in section 3. Results of a set of real implementations are presented in section 4. Section 5 is an analytical observation to MABCVK. A conclusion is drawn in section 6.

## 2. The Scheme

The data to be encrypted is divided in blocks each of L-bits. All these blocks are converted into equivalent set of integers of base-10.Next, two sufficiently large prime integers K1 and K2 are taken as keys satisfying the conditions:

1. K1, K2 being sufficiently bigger than the number or numbers to be encrypted

2. K1 < K2

A variable α with its value being in the range of 0.1 and 0.9 is taken to be used in due course of time to shift the sequence of encrypted set of integers. Sections 2.1 and section 2.2 respectively describe the encryption and the decryption processes with conventionally *Alice* being considered as the sender and *Bob* being considered as the receiver

### 2.1 Encryption

The set of computations to be done in the encryption end for P being a positive integer to be encrypted is as follows:
1. Set K3 = φ (K1) − 1.
2. Set A = Chosen random number from the set of numbers, which when used as quotient to divide K2, leaves P as remainder.
3. Set B = $A^{K3}$ mod K1
4. Set C = Product of B and K2

5. Set D = C$^{K3}$ mod K1
6. Set E = Result of shifting D by α %, which is the encrypted form for P.

The function φ (K1) is a totient function [3]. The totient φ (K1) of a positive integer K1 defines the number of integers less than or equal to K1 that are relatively prime to K1. φ (K1) = K1 − 1, when K1 is a prime number.

## 2.2 Decryption

Based on the set of values for K1, K2, α extracted from the private key, the following set of operations are carried out in the decryption end.

1. Set K3 = φ (K1) − 1.

2. Set D = Result of shifting E by (1- α)%

3. Set B = (D•K2)$^{K3}$ mod K1

4. Set A = B$^{K3}$ mod K1

5. Set P = K2 mod A

The function φ (K1) is a totient function as was described in section 2.1.

## 3. Implementation

This section describes the process of encryption and decryption using a sample tiny plaintext WORLD considering K1 = 263, K2 = 317 and α = 0.1.
Table 1 enlists different intermediate and the final result of implementation following the encryption algorithm stated in section 2.1. Consequently, the cipher text "U≤а¼E" is generated for the taken plaintext

| Table 1 : Implementation of Encryption of the Plaintext "WORLD" | | | | | | | |
|---|---|---|---|---|---|---|---|
| C H A R A C T E R | A S I I V A L U E | Value obtained by following step 2 of the Encryption Algorithm | Value obtained by following step 3 of the Encryption Algorithm | Value obtained by following step 4 of the Encryption Algorithm | Value obtained by following step 5 of the Encryption Algorithm | Character corresponding to value obtained through step 5 of the Encryption Algorithm | Character obtained by following step 6 of the Encryption Algorithm for an assumed value of 0.1 for α |
| W | 87 | 230 | 255 | 80835 | 14 | E | U |
| O | 79 | 119 | 42 | 13314 | 85 | U | ≤ |
| R | 82 | 235 | 216 | 68472 | 243 | ≤ | а |
| L | 76 | 241 | 251 | 79567 | 97 | а | ¼ |
| D | 68 | 83 | 244 | 77348 | 172 | ¼ | E |

A reverse process is followed to employ the decryption algorithm stated in section 2.2.

## 4. Results

The implementation of the encryption / decryption algorithm described in section 2.1 and section 2.2

respectively was accomplished using C language and MIRACL (**M**ulti precision **I**nteger and **R**ational **A**rithmetic **C**/C++ **L**ibrary) library [4] for big integer calculations. The plaintext to be encrypted was read as blocks of 8-bits, a pair of private keys KEY1 and KEY2, each 10-bits long, and a small value **α** was taken for operation on the plaintext data.

Ciphertext data produced after encryption process was written as blocks of 32-bits i.e. size of integer type variable, to avoid overflow, as in some cases 8-bit data after encryption is to yield a value greater than 255. We have encrypted and also successfully decrypted file types .txt, .doc, .docx, .xls, .pdf, .mp4, .ocx, .jpg, .bmp, .dll, .gif, .sys etc.

A comparison of file type, source file size, decrypted file size, encryption time and decryption time is depicted in the table 2.

It is seen from data in Table 2 that the relation between encryption time and file size is linear. The average time, based on tabulated data in table 2, to encrypt 1 KB of data is remarkably found to be 0.028852538 seconds. Similarly, it was noted that the average time required for decryption of 1 KB of data is 0.007198586 seconds.

Each 8-bit source data is written as 32-bit encrypted data, so the encrypted data file size (in bytes) is approximately four times the source data file size (in bytes). Thus it was observed that the proposed cipher carries a schematic overhead of causing data expansion. Having computational complexities, an operational overhead of being marginally slow in execution is observed during its real implementation.

| TABLE 2: file type, source file size, and encryption time | | | | | | | |
|---|---|---|---|---|---|---|---|
| sl no | Source filename | size (in kb) | encryption time (sec) | Encrypted file size (in KB) | Encrypted filename | Decryption Time (sec) | Decrypted File size (in KB) |
| 1 | abc.docx | 47 | 1.516 | 188 | abc.docx.enc | 1.5 | 47 |
| 2 | hawaii.jpg | 18 | 0.64 | 71 | hawaii.jpg.enc | 0.656 | 18 |
| 3 | 3rd_sem2011.docx | 47 | 1.532 | 188 | 3rd_sem2011.docx.enc | 1.5 | 47 |
| 4 | cv.doc | 60 | 1.219 | 238 | cv.doc.enc | 1.235 | 60 |
| 5 | EffectCtrl.ocx | 405 | 11.203 | 1617 | EffectCtrl.ocx.enc | 11.172 | 405 |
| 6 | Flower.bmp | 17 | 0.625 | 68 | Flower.bmp.enc | 0.625 | 17 |
| 7 | hnmaps.dll | 59 | 1.532 | 236 | hnmaps.dll.enc | 1.516 | 59 |
| 8 | internal marks.xls | 85 | 1.781 | 338 | internal marks.xls.enc | 1.719 | 85 |
| 9 | Crypto.zip | 66 | 2.172 | 261 | Crypto.zip.enc | 2.188 | 66 |
| 10 | test.txt | 2 | 0.125 | 7 | test.txt.enc | 0.11 | 2 |
| 11 | Univ_tabulator.pdf | 434 | 13.625 | 1736 | Univ_tabulator.pdf.enc | 13.688 | 434 |
| 12 | Vivekananda.gif | 25 | 0.875 | 100 | Vivekananda.gif.enc | 0.875 | 25 |
| 13 | BuzzingBee.wav | 144 | 4.829 | 573 | BuzzingBee.wav.enc | 4.765 | 144 |
| 14 | clock.avi | 81 | 2.578 | 324 | clock.avi.enc | 2.484 | 81 |
| 15 | EffectCtrl.ocx | 405 | 11.219 | 1617 | EffectCtrl.ocx.enc | 11.266 | 405 |
| 16 | Gone Fishing.bmp | 17 | 0.625 | 68 | Gone Fishing.bmp.enc | 0.641 | 17 |
| 17 | Greenstone.bmp | 26 | 0.875 | 104 | Greenstone.bmp.enc | 0.844 | 26 |
| 18 | handsafe.reg | 1 | 0.031 | 3 | handsafe.reg.enc | 0.031 | 1 |
| 19 | himem.sys | 5 | 0.234 | 19 | himem.sys.enc | 0.234 | 5 |
| 20 | IE7Eula.rtf | 73 | 2.469 | 292 | IE7Eula.rtf.enc | 2.485 | 73 |
| 21 | jysew.exe | 151 | 3.156 | 602 | jysew.exe.enc | 3.14 | 151 |
| 22 | Keyboard.drv | 2 | 0.094 | 8 | Keyboard.drv.enc | 0.109 | 2 |
| 23 | Mscomm32.ocx | 102 | 2.828 | 406 | Mscomm32.ocx.enc | 2.765 | 102 |
| 24 | normnfd.nls | 39 | 1.063 | 154 | normnfd.nls.enc | 1.031 | 39 |
| 25 | notepad.exe | 68 | 1.797 | 270 | notepad.exe.enc | 1.734 | 68 |
| 26 | PreConvertLite.dll | 103 | 2.969 | 412 | PreConvertLite.dll.enc | 2.953 | 103 |

Figure 1 shows plotting of source file size (in KB) and encryption time (in Sec).

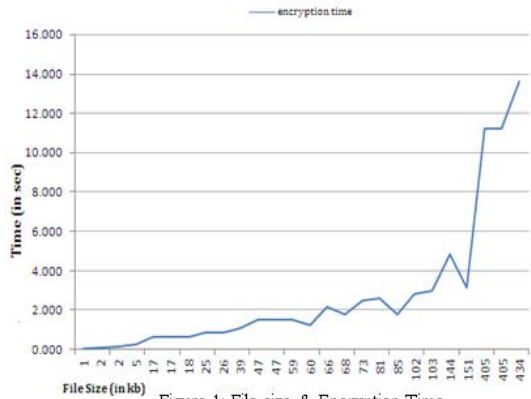

Figure 1: File size & Encryption Time

Figure 2 shows plotting of file size and decryption time.

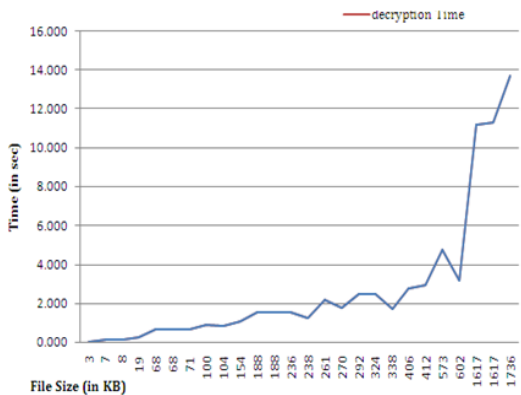

Figure 2: File size & Decryption Time

The encryption and decryption process being almost symmetrical, the encryption time for any specific file is nearly equal to the time required to decrypt it. This is validated from the graph plotting shown in figure 1 and figure 2.

## 5. Analysis

Although small keys were used to tabulate results of encryption using MABCVK, it can be used with much larger key-spaces with the support of appropriate computing platform for large exponential computations.

The substitution of plaintext data with random integers generates different ciphers for the same set of keys. If the number of integers to be encrypted is N and if there can be at least one substitution for each then there can be at least N different cipher texts generated using the same key sets. This property makes frequency analysis attacks difficult.

The effective key length is equal to sum of length of KEY1 and KEY2. Brute-force search will require searching entire search space of $2^{(sum\ of\ length\ of\ KEY1\ \&\ KEY2)}$.

The average time required for such extensive search is given in table 3 [5].

| TABLE 3: Average Required Time for Exhaustive Key Search | | | |
|---|---|---|---|
| Key Size(Bits) | Number of Alternative Keys | Time Required at 1 Encryption /µs | Time Required at $10^6$ encryptions /µs |
| 56 | $2^{56} = 7.2 \times 10^{16}$ | $2^{55}$ µs =1142 years | 10.01 hours |
| 128 | $2^{128} = 3.4 \times 10^{38}$ | $2^{127}$ µs = 5.4 X $10^{24}$ years | 5.4 X $10^{18}$ years |
| 168 | $2^{168} = 3.7 \times 10^{50}$ | $2^{167}$ µs = 5.9 X $10^{36}$ years | 5.9 X $10^{30}$ years |
| 26 characters | $26! = 4 \times 10^{26}$ | $2 \times 10^{26}$ µs = 6.4$X10^{12}$ years | 6.4 X $10^6$ years |

Data expansion noted during encryption is expected as the proposed cipher uses exponentiation. We can adopt to write encrypted data as bit stream and device an appropriate padding scheme to identify separate blocks to reduce the encrypted file size considerably. The computational complexity of calculating large integer powers of integers can be simplified by adopting faster methods like exponentiation by squaring, square and multiply or binary exponentiation, which leads to reduction of both encryption and decryption time. Encryption process can be augmented to be faster if read block size (in bits) is increased but it should be less than the size of key (in bits). The proposed scheme restricts use of the read block size greater than the key size.

## 6. Conclusion

The proposed MABCVK is flexible enough for implementation with larger keys. The substitution of source data with random integers produces different sets of encrypted data for the same private key pairs, which, in turn, makes the task of cryptanalysis a challenging one. Encryption and decryption using MABCVK require almost same execution time. If a pair of 128-bit keys is used for encryption and if brute-force attack is applied for decryption then at least all the primes in the range of integers from 1 to $2^{128}$ are to be tested as a probable key. There are $2^{128}$ / Log ($2^{128}$) = 383534127545935 x $10^{22}$ prime numbers in the range of integers from 1 **to $2^{128}$**. It is observed that if one trillion of these numbers per second can be checked, more than 121,617,874,031,562 x **$10^3$** years is required to check all such primes. The requirement of time as shown in table 3, to calculate the keys makes it reasonably non-vulnerable against cryptanalytic attacks. As a part of the future scope of the work, it is also intended to formulate and implement an enhanced scheme that will use a carrier image to have a secure transmission of the private keys.

Abhijit Chowdhury, MCA, Assistant Professor, Department of Computer Applications – NSHM College of Management & Technology, Durgapur, West Bengal, INDIA. Member of IAENG (International Association of Engineers, http://www.iaeng.org). Research Area of interest is cryptography, network security.

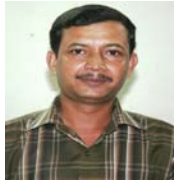

Angshu Kumar Sinha, M.Sc.-Mathematics, Department of Computer Applications – NSHM College of Management & Technology, Durgapur, West Bengal, INDIA.

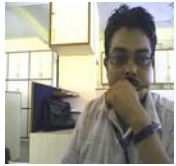

Dr. Saurabh Dutta, Ph.D. (Computer Science), Dr. B. C. Roy Engineering College, Durgapur-713206, West Bengal, INDIA.

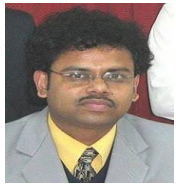